\documentclass[prl,twocolumn,eqsecnum,floatfix,superscriptaddress,preprintnumbers,amsmath,amssymb,nofootinbib,aps]{revtex4-2}

\usepackage{hyperref}
\usepackage{graphicx}
\usepackage{dcolumn}
\usepackage{bm}
\usepackage{amsmath}
\usepackage{xcolor}
\usepackage{soul}

\usepackage{graphicx}
\usepackage{dcolumn}
\usepackage{bm}
\def\lesssim{\ \raise.3ex\hbox{$<$}\kern-0.8em\lower.7ex\hbox{$\sim$}\ }
\def\gesim{\ \raise.3ex\hbox{$>$}\kern-0.8em\lower.7ex\hbox{$\sim$}\ }

\begin{document}
\preprint{RIKEN-iTHEMS-Report-26}
\title{Nonequilibrium Andreev transport at the QGP-2SC interface}
\author{Tingyu Zhang}
\affiliation{Department of Physics, State Key Laboratory of Optical Quantum Materials, and Hong Kong Institute of Quantum Science and Technology, University of Hong Kong, Hong Kong, China}

\author{Honoka Hoshino}
\affiliation{Department of Physics, Saga University, Saga 840-8502, Japan}

\author{Hiroyuki Tajima}
\affiliation{Department of Physics, Graduate School of Science, The University of Tokyo,
    Tokyo 113-0033, Japan}
\affiliation{RIKEN Nishina Center, Wako 351-0198, Japan}
\affiliation{Quark Nuclear Science Institute, The University of Tokyo, Tokyo 113-0033, Japan}

\author{Motoi Tachibana}
\affiliation{Department of Physics, Saga University, Saga 840-8502, Japan}
\affiliation{Center for Theoretical Physics, Khazar University, 41 Mehseti Street, Baku, AZ1096, Azerbaijan}

\author {Mariusz Sadzikowski}
\affiliation{Jagiellonian University, Institute of Theoretical Physics, Kraków, Poland}

\date{\today}
\begin{abstract}

We discuss a nonequilibrium Andreev reflection at an interface between quark-gluon plasma (QGP) and two-flavor color superconducting (2SC) quark matter. Based on the Schwinger-Keldysh framework and a relativistic tunneling model, we evaluate the momentum-resolved tunneling current generated by a chemical-potential bias between the QGP and 2SC phases. We find that the Andreev reflection appears as at the fourth order of the tunneling strength, in which an incident quark in QGP is converted into a reflected hole, while a Cooper pair is injected into the 2SC condensate. We show that the Andreev reflection is enhanced when the bias becomes comparable to the gap and is suppressed in the supergap region, which is similar to that in superconducting materials.
The present formulation provides a field-theoretical pathway to strongly-correlated transport across dense-matter interfaces relevant to nonequilibrium dynamics in compact stars.
\end{abstract}

\maketitle

\textit{Introduction}.---The physics of dense quantum chromodynamics (QCD) matter is central to compact-star structure and dynamics.  At sufficiently high baryon density, deconfined quark matter may appear in the inner region of massive neutron stars, while the surrounding lower-density region may contain hadronic matter with nucleon superfluidity \cite{RevModPhys.75.607,Baym_2018,haskell2018superfluidity}. At even higher density, attractive quark-quark correlations near the Fermi surface can produce color superconductivity \cite{BARROIS1977390,frautschi1980asymptotic,RevModPhys.80.1455}.  Interfaces among hadronic matter, normal quark matter, and color-superconducting quark matter may therefore play a nontrivial role in transport, dissipation, and relaxation processes in compact stars.

One of the characteristic interfacial processes associated with superconductivity is an Andreev reflection, in which an incident particle is converted into a reflected hole while a Cooper pair enters the condensate~\cite{osti_4071988,RevModPhys.80.1337}. In addition to condensed matter, analogous processes have been discussed in ultracold atomic gases~\cite{PhysRevLett.100.110404,PhysRevLett.102.180405,PhysRevResearch.4.023231,zhang2023dominant} and relativistic dense quark matter, including interfaces involving two-flavor color superconducting (2SC) and color-flavor locked (CFL) phases \cite{Sadzikowski:2002ib,PhysRevD.66.045024,sadzikowski2002andreev}. These studies employ a relativistic Bogoliubov-de Gennes (BdG) scattering approach, in which quasiparticle wave functions are matched across a sharp interface. In contrast to ordinary metallic systems, however, dense QCD matter involves color, flavor, and relativistic particle-hole, particle-antiparticle excitations, and possible tunneling processes involving multi-particle cluster formations, that is beyond simple wave-function matching.

A promising way to avoid this limitation is to work directly with nonequilibrium Green's functions.
The Schwinger-Keldysh approach~\cite{schwinger1961brownian,Keldysh,PhysRev.124.287,PhysRev.127.1391,Rammer_2007,Stefanucci} combined with a tunneling Hamiltonian gives a systematic perturbative expansion of interface currents in terms of contour-ordered correlation functions and has been widely used in condensed matter~\cite{RevModPhys.58.323,PhysRevB.50.5528,PhysRevLett.68.2512,BELZIG19991251,wang2014nonequilibrium,PhysRevLett.120.037201,PhysRevB.99.144411,PhysRevLett.124.166803,kawamura2026nonequilibrium} and cold atomic systems~\cite{Sieberer_2016,PhysRevA.95.013623,PhysRevLett.118.105303,furutani2020strong,PhysRevA.106.033310,tajima2023nonequilibrium,PhysRevB.108.155303,PhysRevApplied.21.L031001,PhysRevB.110.064512,kawamura2024non,PhysRevLett.133.163402}. Moreover, such a formalism has been used to study the tunneling transport between color superconducting quark matter and hadron superconducting matter~\cite{zhang2025}.
However, the Lorentz-covariant Dirac structure and the explicit color-flavor structure of quark Cooper pairs are not retained in Ref.~\cite{zhang2025}. These properties are crucial for understanding color-superconducting transport in massive neutron stars. 

In this Letter, we discuss nonequilibrium Andreev transport between quark-gluon plasma (QGP) and 2SC quark matter, by overcoming the limitations and difficulties in previous work, namely, with the use of the Schwinger-Keldysh approach without resorting to the wave-function matching and the nonrelativistic approximation. 
Another important advantage of our approach is to directly access macroscopic observables such as tunneling currents and density relaxations. 
Considering the tunneling rate of the momentum distribution, we elucidate microscopic features of particle and antiparticle conversions during the Andreev transport process, which would be relevant to astrophysical phenomena in massive neutron stars.

\textit{Model}.---Throughout this work we take the unit of $\hbar=k_{\rm B}=1$. We apply a non-equilibrium quantum field theory to describe a relativistic interface between normal QGP and 2SC quark matter. In the 2SC phase, the up and down quarks form a Cooper pair in a spin-zero, color-antitriplet, and flavor-singlet channel. Equivalently, in a suitable color basis, the red and green quarks participate in the condensate, whereas the blue component is not paired by the 2SC gap~\cite{PhysRevD.66.094007,alford2001color,RevModPhys.80.1455}. 
The bulk subsystems of baryon matter and quark matter are assumed to be in the thermodynamic limit, and their volumes are taken to be unity. 
Considering a planar interface that separates a normal QGP region and a 2SC quark-matter region, the total effective Hamiltonian is written as
\begin{align}
    H=H_{\rm QGP}+H_{\rm 2SC}+H_{\rm T}.
\end{align}
Here $H_{\rm QGP}$ describes quarks in the normal phase,
\begin{equation}\label{eq:HQGP}
  H_{\rm QGP}=\sum_{a,i,\bm{k}} q^{i\dagger}_{\bm{k}a}
  \left(\bm{\alpha}\cdot\bm{k}+m\gamma_0-\mu_{\rm QGP}\right)
  q^i_{\bm{k}a},
\end{equation}
where $a=r,g,b$ denotes color and $i=u,d$ denotes flavor. $q^i_{\bm{k}a}$ is the QGP quark operator with momentum $\bm{k}$, flavor $i$, and color $a$, carrying a suppressed Dirac spinor index. Therefore, the matrix $\bm{\alpha}\cdot\bm{k}+m\gamma_0$ acts in the Dirac space. $m$ is the quark mass, $\mu_{\rm QGP}$ is the quark chemical potential in the QGP phase, and $\bm{\alpha}$ and $\gamma_0$ are the usual Dirac matrices.

The 2SC side is modeled by the mean-field Hamiltonian
\begin{align}\label{eq:H2SC}
  H_{\rm 2SC}={}&\sum_{a,i,\bm{k}} Q^{i\dagger}_{\bm{k}a}
  \left(\bm{\alpha}\cdot\bm{k}+m\gamma^0-\mu_{\rm 2SC}\right)
  Q^i_{\bm{k}a} \nonumber\\
  &+\frac{\Delta}{2}\sum_{a,b,i,j,\bm{k}}
  Q^{i\dagger}_{\bm{k}a}\mathcal{M}^{ij}_{ab}
  Q^{j\dagger}_{-\bm{k}b}
  \nonumber\\
  &-\frac{\Delta^*}{2}\sum_{a,b,i,j,\bm{k}}
  Q^i_{-\bm{k}a}\mathcal{M}^{ij}_{ab}Q^j_{\bm{k}b}
  +{\rm const.},
\end{align}
where $Q^i_{\bm{k}a}$ is the quark operator in the 2SC phase and $\Delta$ is the value of the 2SC gap parameter. 
The 2SC pairing matrix is $\mathcal{M}^{ij}_{ab}=(\tau_2)^{ij}t_{ab}C\gamma^5$, where $\tau_2$ is the antisymmetric Pauli matrix in the flavor space, $t$ is an antisymmetric Gell-Mann matrix in the color space, and $C$ is the charge-conjugation matrix.  

The interface is described by an effective single-quark tunneling Hamiltonian
\begin{equation}\label{eq:HT}
  H_{\rm T}=\sum_{\bm{k},\bm{k}',a,b,i,j}
  \mathcal{T}_{\bm{k}\bm{k}',ab}^{ij}
  Q^{i\dagger}_{\bm{k}a}\mathcal{M}^{ij}_{ab}q^j_{\bm{k}'b}+\mathrm{h.c.},
\end{equation}
where $\mathcal{T}_{\bm{k}\bm{k}',ab}^{ij}$ is the tunneling coupling strength. Note that Eq.~(\ref{eq:HT}) is an effective low-energy tunneling Hamiltonian for quasiparticles near the QGP–2SC interface. The matrix $\mathcal{M}_{ab}^{ij}$
projects the interfacial coupling onto the color-flavor-Dirac channel associated with the 2SC condensate, rather than representing a general microscopic QCD interaction. The value of $\mathcal{T}_{\bm{k}\bm{k}',ab}^{ij}$ should depend on the shape, thickness, and roughness of the interface. Below, unless otherwise specified, we take $\mathcal{T}_{\bm{k}\bm{k}',ab}^{ij}=\mathcal{T}$ to be real and momentum independent to expose the qualitative structure of the tunneling current.

\textit{Keldysh formulation of tunneling current}.---We define the quark momentum-occupation operator on the QGP side for a given channel $\bm{k},a,i$ as $N_{\bm{k}ai,\rm QGP}=q^{i\dagger}_{\bm{k}a}q^i_{\bm{k}a}$ with no sum of repeated indices,
where the Dirac spinor index for $q^\dagger$ and $q$ is implicitly contracted.
The current flowing out of QGP into the 2SC region is then defined by the rate of decrease of the momentum-occupation operator $\hat I_{\bm{k}ai}=i [N_{\bm{k}ai,\rm QGP},H_{\rm T}]$.
With the tunneling Hamiltonian in Eq.~\eqref{eq:HT}, this becomes
\begin{equation}\label{eq:Ioperator}
  \hat I_{\bm{k}ai}=-i\sum_{\bm{k}',b,j}
  \mathcal{T}_{\bm{k}'\bm{k},ba}^{ji} Q^{j\dagger}_{\bm{k}'b}\mathcal{M}^{ji}_{ba}q^i_{\bm{k}a}
  +\mathrm{h.c.}.
\end{equation}
The sign convention is such that a positive current corresponds to the depletion of the QGP occupation in the specified mode.

The nonequilibrium expectation value of the current is evaluated by expanding in terms of the tunneling Hamitonian $H_{\rm T}$:
\begin{align}\label{eq:Kexpansion}
  I_{\bm{k}ai}(t,t')={}&
  \sum_{n=0}^{\infty}\frac{(-i)^n}{n!}
  \int_C d t_1\cdots \int_C d t_n
  \nonumber\\
  \times&
  \left\langle {\rm T}_C \hat I_{\bm{k}ai}(t,t')
  H_{\rm T}(t_1)\cdots H_{\rm T}(t_n)\right\rangle_0,
\end{align}
where $\langle\cdots\rangle_0$ denotes the average with respect to the decoupled QGP and 2SC Hamiltonians. The contour $C$ denotes the Keldysh contour consisting of a forward branch along the real time axis and a backward one, while $t$ and $t'$ denote the time points locating on the backward and forward branches, respectively. 
The physical current is obtained by taking the equal-time limit $t'\to t$ after applying the Langreth rules.

\begin{figure}[t]
    \centering
    \includegraphics[width=8.5cm]{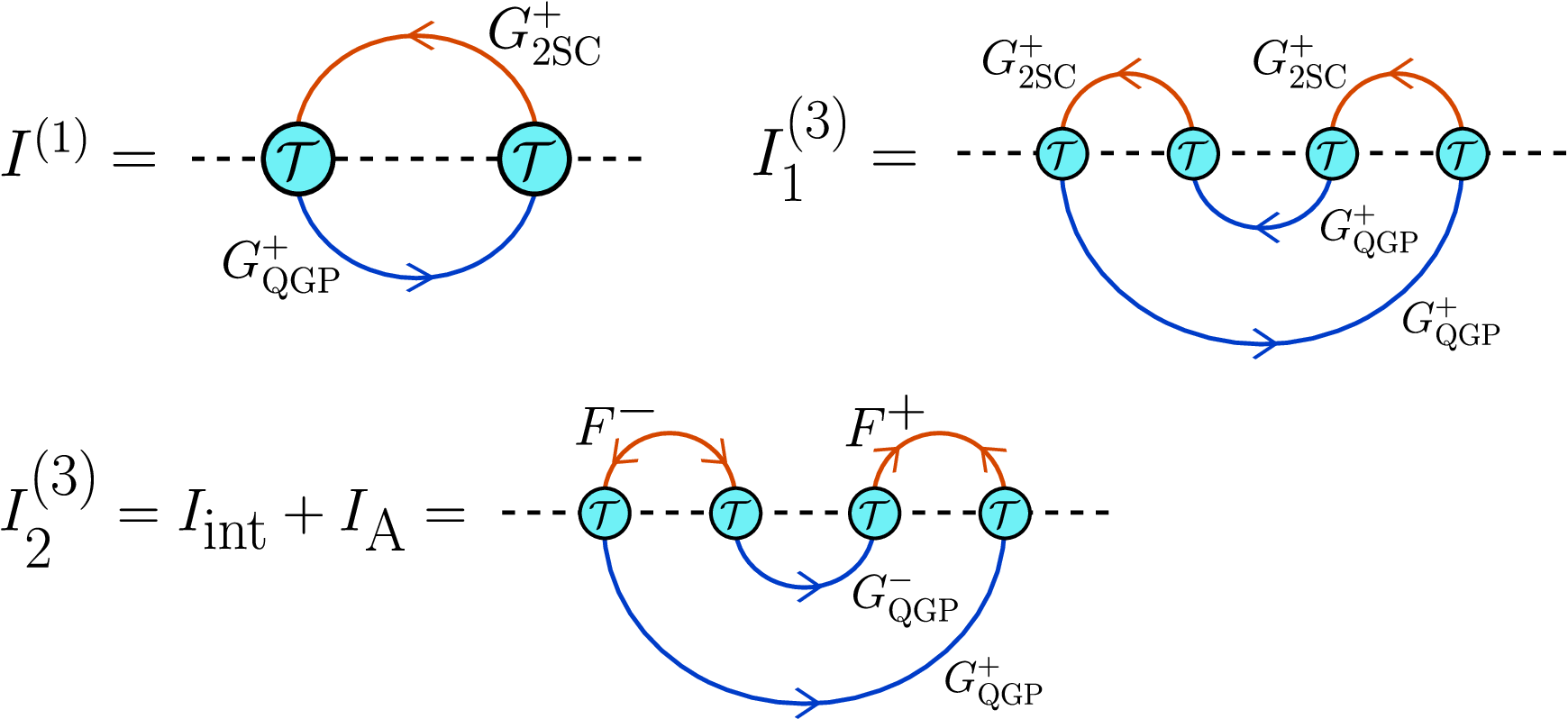}
    \caption{Diagrammatic representation of the first-order contribution $I^{(1)}$, and third-order contribution $I_1^{(3)}$ and $I_2^{(3)}$ to the tunneling current. The blue and red solid lines denote quark propagators in QGP and 2SC phases, respectively, within the Nambu-Gor'kov representation. The circle represents the tunneling coupling $\mathcal{T}$.}
    \label{fig:diagrams}
\end{figure}
We assume that the QGP and 2SC regions act as two steady reservoirs characterized by chemical potentials $\mu_{\rm QGP}$ and $\mu_{\rm 2SC}$, respectively. In Heisenberg representation, the operators evolve with grand-canonical Hamiltonian rather than the bare Hamiltonian. Therefore, operators in Eq.~(\ref{eq:Kexpansion}) should be transformed as $q^i_{\bm{k}a}(t)\rightarrow e^{-i\mu_{\rm QGP}t}q^i_{\bm{k}a}(t)$ and $Q^i_{\bm{k}a}(t)\rightarrow e^{-i\mu_{\rm 2SC}t}Q^i_{\bm{k}a}(t)$.

Using these, we find the leading nonvanishing term appearing as the first order in the contour expansion. 
We introduce the Green's function in the Nambu-Gor'kov space~\cite{Supplement}:
\begin{equation}\label{eq:NGmatrix}
  i \mathcal{G}^{ij}_{\alpha,\bm{k}ab}(t,t')=
  \begin{pmatrix}
  G^{ij+}_{\alpha,\bm{k}ab}(t,t') & F^{ij-}_{\alpha,\bm{k}ab}(t,t')\\
  F^{ij+}_{\alpha,\bm{k}ab}(t,t') & G^{ij-}_{\alpha,\bm{k}ab}(t,t')
  \end{pmatrix}.
\end{equation} 
In the QGP region, the anomalous components vanish,
$F^{ij,\pm}_{\rm QGP,\bm{k},ab}=0$, because there is no pairing condensate. Moreover, normal components are diagonal in color and flavor space, $G^{ij\pm}_{{\rm QGP},\bm{k}ab}=\delta^{ij}\delta_{ab}
G^{\pm}_{\rm QGP,\bm{k}}$. Therefore, in the following formulas, we only label one color index and one flavor index for the Green's functions. 
Wick's theorem and Langreth rules give a frequency representation of the leading-order contribution in terms of normal Green's functions:
\begin{align}\label{eq:I1spectral}
  I^{(1)}_{\bm{k}ai}=&4\sum_{\bm{k}',b,j}
  \big|\mathcal{T}\big|^2
  \int\frac{d\omega}{2\pi}\
  {\rm Tr}\big[
  \mathrm{Im}\,G^{i+}_{\rm QGP,\bm{k}a}(\omega-\Delta\mu)\nonumber\\
  &\times\mathrm{Im}\,G^{j+}_{\rm 2SC,\bm{k}'b}(\omega)
  \big][f(\omega-\Delta \mu)-f(\omega)],
\end{align}
where $\Delta\mu=\mu_{\rm QGP}-\mu_{\rm 2SC}$.
This is the ordinary quasiparticle tunneling current, and its diagrammatic representation is shown in Fig.~\ref{fig:diagrams}.

\textit{Pair correlation tunneling and Andreev reflection}.---The second order in the contour expansion vanishes by field parity. The next nonzero contribution is the third order in $H_{\rm T}$, which includes two classes. The first class contains only normal propagators and gives a higher-order quasiparticle correction, denoted by $I_1^{(3)}$
shown diagrammatically in Fig.~\ref{fig:diagrams}.

The second class contains anomalous 2SC propagators and is responsible for pair-correlated transport. In 2SC pairing structure, the matrix in flavor-color space is given by
\begin{align}\label{eq:tau2t}
    \tau_2 t=\left(\begin{array}{cc}
0 & \ -i \\
i & \ 0
\end{array}\right) \otimes\left(\begin{array}{ccc}
0 & \ -i & \ 0 \\
i & \ 0 & \ 0 \\
0 & \ 0 & \ 0
\end{array}\right),
\end{align}
which is a symmetric matrix with nonzero elements $(\tau_2 t)_{rg}^{ud}=(\tau_2 t)_{gr}^{du}=-1$ and $(\tau_2 t)_{gr}^{ud}=(\tau_2 t)_{rg}^{du}=1$, meaning that the pairing occurs only between $ru$ and $gd$ (red up and green down), as well as between $rd$ and $gu$ (red down and green up).

After Wick contraction for all possible arrangements of the color and flavor indices, we obtain the rest part of the third expansion term containing anomalous 2SC propagators as $I^{(3)}_{2,\bm{k}}=I_{\mathrm{int},\bm{k}}+I_{{\rm A},\bm{k}}$, where 
\begin{widetext}
\begin{align}\label{eq:Iint}
    I_{{\rm int},\bm{k}ai}=&16\sum_{\bm{k}'\bm{k}_1\bm{k}_1',bcd,jkl} \mathcal{T}^4\int \frac{d\omega}{2\pi}\operatorname{Tr}\big[\operatorname{Im}G^{i+}_{{\rm QGP},\bm{k}a}(\omega-\Delta\mu)\operatorname{Im}F^{j-}_{\bm{k}'b}(\omega)\operatorname{Im}G^{k-}_{{\rm QGP},\bm{k}'_1c}(\omega+\Delta\mu)\operatorname{Im}F^{l+}_{\bm{k}_1d}(\omega)\cr
    &-\operatorname{Im}G^{i+}_{{\rm QGP},\bm{k}a}(\omega-\Delta\mu)\operatorname{Im}F^{j-}_{\bm{k}'b}(\omega)\operatorname{Re}G^{k-}_{{\rm QGP},\bm{k}'_1c}(\omega+\Delta\mu)\operatorname{Re}F^{l+}_{\bm{k}_1d}(\omega)\big]
    [f(\omega)-f(\omega-\Delta\mu)],
\end{align}
\begin{align}\label{eq:IA}
    I_{{\rm A},\bm{k}ai}=&8\sum_{\bm{k}'\bm{k}_1\bm{k}_1',bcd,jkl} \mathcal{T}^4\int \frac{d\omega}{2\pi}
    \operatorname{Tr}\big[\operatorname{Im}G^{i+}_{{\rm QGP},\bm{k}a}(\omega-\Delta\mu)\operatorname{Im}F^{j-}_{\bm{k}'b}(\omega)\operatorname{Im}G^{k-}_{{\rm QGP},\bm{k}'_1c}(\omega+\Delta\mu)\operatorname{Im}F^{l+}_{\bm{k}_1d}(\omega)\cr
    &+\operatorname{Im}G^{i+}_{{\rm QGP},\bm{k}a}(\omega-\Delta\mu)\operatorname{Re}F^{j-}_{\bm{k}'b}(\omega)\operatorname{Im}G^{k-}_{{\rm QGP},\bm{k}'_1c}(\omega+\Delta\mu)\operatorname{Re}F^{l+}_{\bm{k}_1d}(\omega)\big][f(\omega-\Delta\mu)-f(\omega+\Delta\mu)].
\end{align}
\end{widetext}
Eq.~(\ref{eq:Iint}) represents the intermediate current describing a single-quark tunneling process dressed by virtual anomalous propagation in the 2SC region, while Eq.~(\ref{eq:IA}) is the Andreev current describing the conversion of a QGP quark into a QGP hole together with injection of a diquark Cooper pair into the 2SC condensate. The diagrammatic representation is shown in Fig.~\ref{fig:diagrams} as the third.


\begin{figure}[t]
    \centering
    \includegraphics[width=8.6cm]{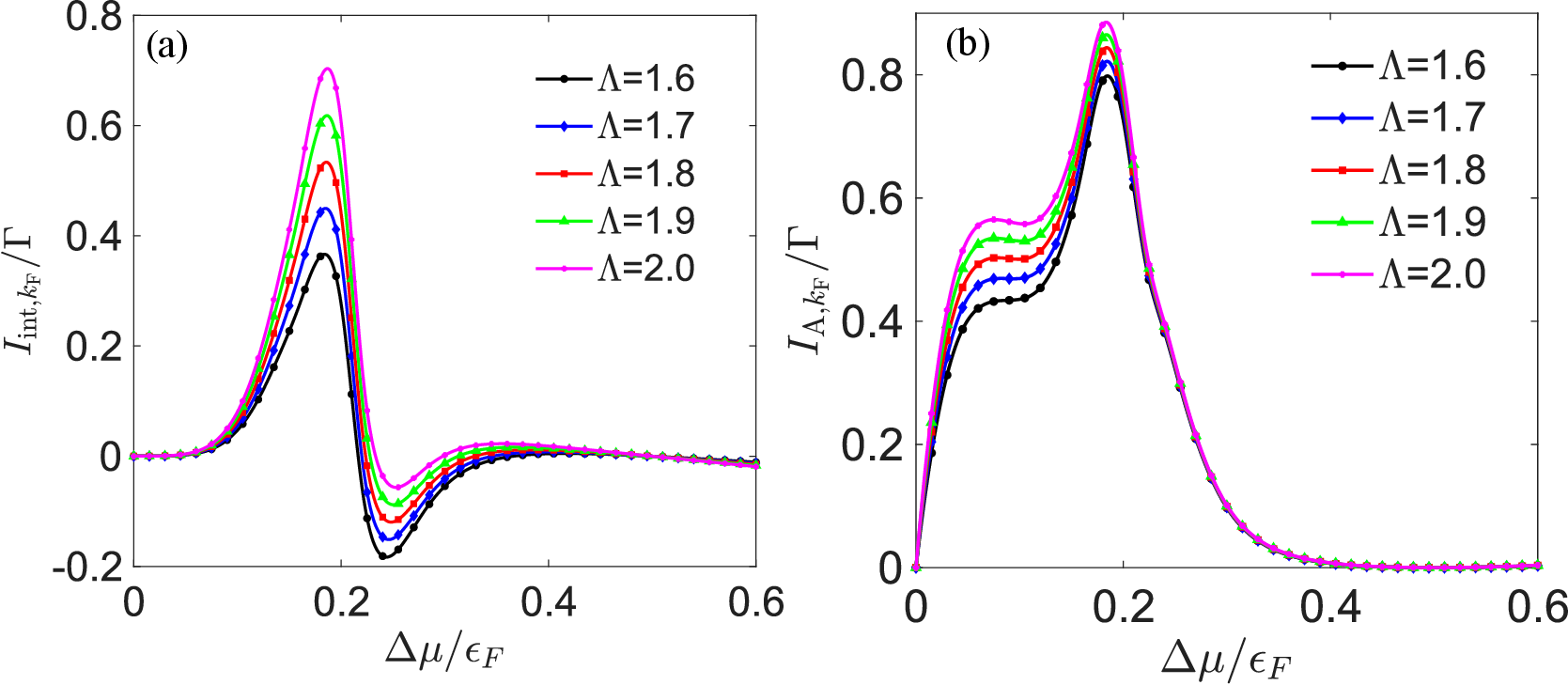}
    \caption{(a) The intermediate current $I_{{\rm int},k_{\rm F}ai}$ and (b) the Andreev current $I_{{\rm A},k_{\rm F}ai}$ for red or green charge as a function of chemical potential bias between QGP and 2SC phases with different momentum cutoff $\Lambda$. Subscripts $a$ and $i$ in the figures are omitted since the currents are independent of color and flavor charges. The value of 2sc gap is taken as $|\Delta|/\epsilon_{\rm F}=0.2$. The temperature is taken as $T/T_{\rm F}\simeq 0.05$ ($T\simeq 25$~MeV). $\Gamma=27\mathcal{T}^4k_{\rm F}^6/(2\pi^3)$ is the normalizing factor.
    }
    \label{fig:I-bias}
\end{figure}
For an explicit evaluation, we use massless relativistic propagators (see the Supplemental Material). We discuss the qualitative behavior of the pair-correlated currents $I_{\rm int}$ and $I_{\rm A}$.
The momentum integrals are regularized by an ultraviolet cutoff $\Lambda$. This momentum cutoff represents the scale in which the low-energy effective tunneling model for quasiparticles is valid. Denoting the QGP Fermi momentum and Fermi energy as $k_{\rm F}$ and $\epsilon_{\rm F}\simeq\mu_{\rm QGP}$, Fig.~\ref{fig:I-bias} shows $I_{{\rm int},k_{\rm F},a,i}$ and $I_{{\rm A},k_{\rm F},a,i}$ as functions of the chemical potential bias $\Delta\mu/\epsilon_{\rm F}$. Taking a typical quark Fermi energy $\epsilon_{\rm F}\simeq \mu_{\rm QGP}\simeq 500~\mathrm{MeV}$ and a representative 2SC gap $|\Delta|\simeq100~\mathrm{MeV}$ gives $|\Delta|/\epsilon_{\rm F}\simeq0.2$~\cite{PhysRevD.74.074020,PhysRevD.82.045010}. Both $I_{\mathrm{int}}$ and $I_A$ are enhanced when the bias becomes comparable to the 2SC gap, $\Delta \mu\simeq |\Delta|$, 
due to the overlap between the QGP spectral branch and the anomalous spectral weight of the 2SC quasiparticles. The intermediate current $I_{\rm int}$ corresponds to a single-quark tunneling process dressed by virtual diquark correlations in the 2SC phase. Physically, the quark couples to an intermediate anomalous pair in 2SC phase before returning to the normal quasiparticle state, which may lead to an interference between quasiparticle and pair tunneling amplitudes and thus can be negative. The Andreev reflection, by contrast, is associated directly with pair transfer and is strongly suppressed in the supergap region, which is in analogy with the ordinary N-S junction in condensed matter.

\begin{figure}[t]
    \centering
    \includegraphics[width=8.6cm]{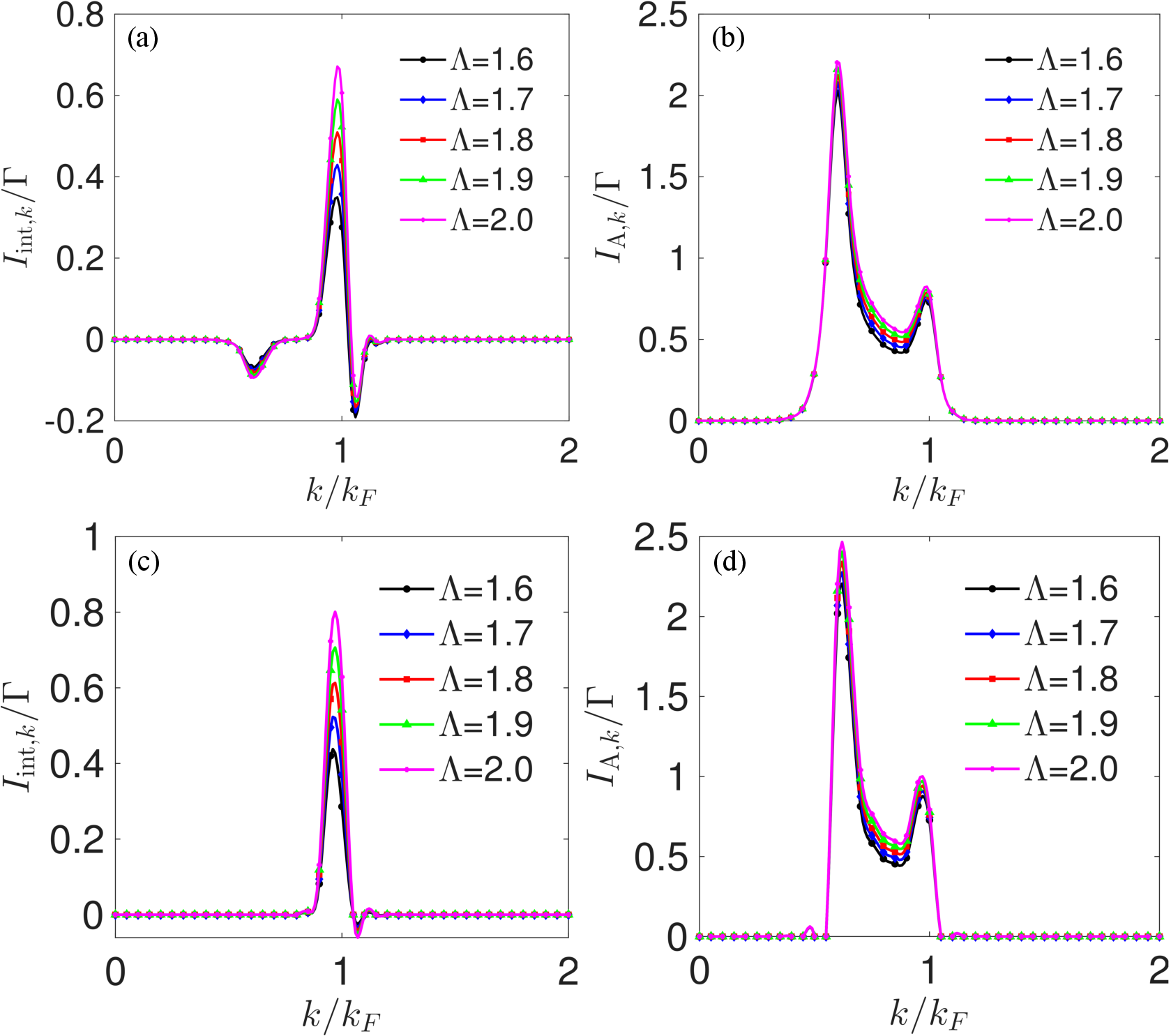}
    \caption{(a) The Intermediate current $I_{{\rm int},kai}$ and (b) the Andreev current $I_{{\rm A},kai}$ as a function of momentum $k/k_{\rm F}$ with different momentum cutoff $\Lambda$ at $T/T_{\rm F}\simeq 0.05$ ($T\simeq 25$~MeV). (c) and (d) respectively shows the momentum dependence of $I_{\rm int}$ and $I_{\rm A}$ at $T\rightarrow 0$. The values of chemical potential bias and the 2SC gap are taken as $\Delta\mu/\epsilon_{\rm F}=|\Delta|/\epsilon_{\rm F}=0.2$. $\Gamma$ is the normalizing factor.}
    \label{fig:I-k}
\end{figure}
Fig.~\ref{fig:I-k} shows the k-dependence of $I_{\rm int}$ and $I_{\rm A}$, where we take $\Delta\mu/\epsilon_{\rm F}=|\Delta|/\epsilon_{\rm F}=0.2$. The peaks in panels~(a) and (b) come from resonances between the QGP spectral function and the anomalous 2SC spectral branches. According to form of the Green's functions, we know that the QGP spectral function peaks at $\omega-\Delta\mu+\mu_{\rm QGP}-k\simeq0$ and the 2SC anomalous spectral function peaks at $\omega\simeq\pm E_{\bm{k}'}$. For momenta near the 2SC Fermi surface, $E_{\bm{k}'}\simeq|\Delta|$. Therefore, resonances occur when 
\begin{align}
    \Delta\mu-\mu_{\rm QGP}+k\simeq \pm|\Delta|.
\end{align}
That is, $k/k_{\rm F}\simeq1.0$ and $0.6$, while the opposite sign of $I_{\rm int}$ appearing at $k/k_{\rm F}\simeq0.6$ arises from the opposite sign of the negative frequency branch in $\operatorname{Im}F^-(\omega)$, which peaks at $\omega\simeq-|\Delta|$. Physically, the QGP quark can tunnel most efficiently when its bias-shifted energy matches an available pair-correlated excitation channel in the 2SC phase. Panels (c) and (d) show their momentum dependence at low temperature limit. When $T\rightarrow 0$, the Fermi functions become sharp step functions. As a result, according to Eq.~(\ref{eq:Iint}), $I_{\rm int}$ is restricted to the positive-energy bias window $0<\omega<\Delta\mu$, and the negative-frequency branch corresponding to $\omega\simeq -|\Delta|$ is suppressed, indicating that the hole-like quasiparticle is Pauli blocked and does not make a net contribution to the intermediate tunneling process. By contrast, the Andreev current is governed by the symmetric window $-\Delta\mu<\omega<\Delta\mu$ so the hole-like branch remains active. This reflects the fact that the hole conversion is intrinsic to Andreev reflection even at $T\simeq 0$.

\begin{figure}[t]
    \centering
    \includegraphics[width=8.6cm]{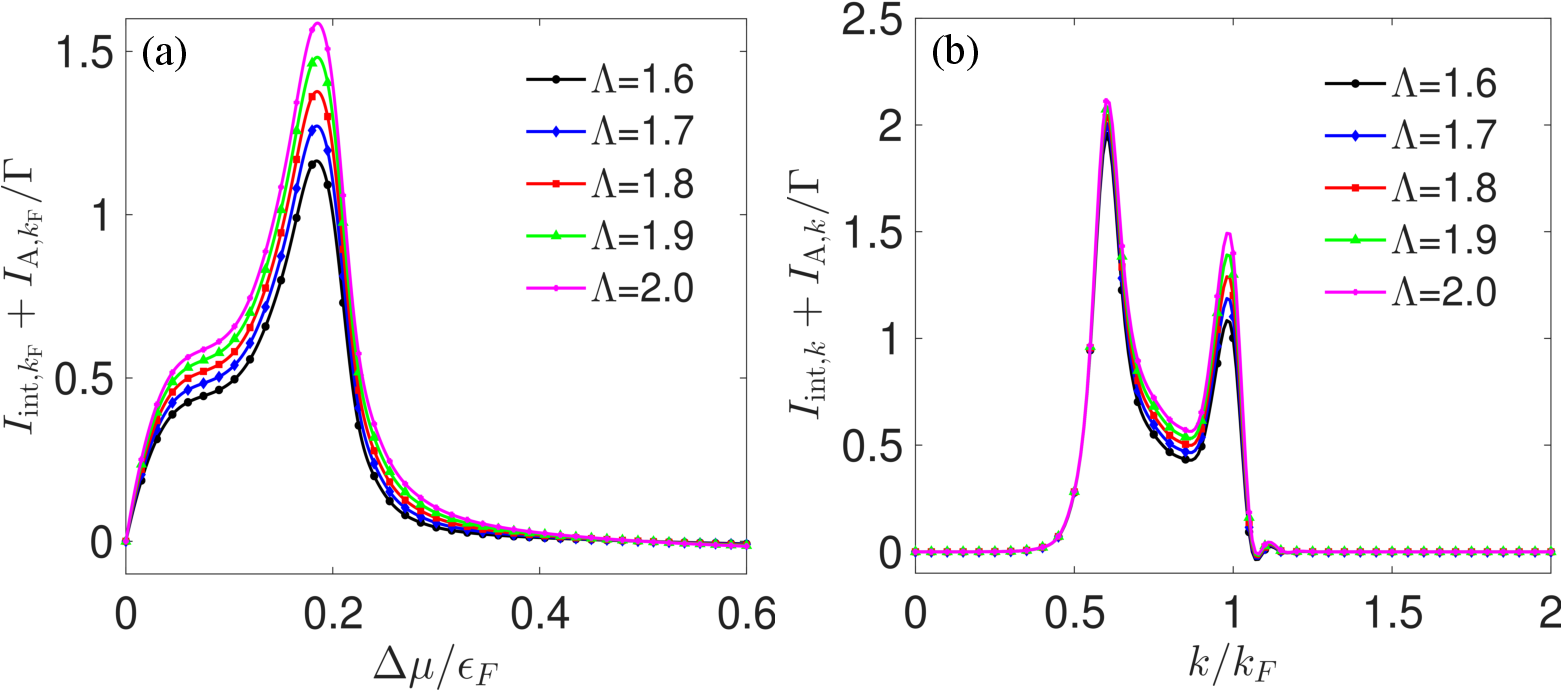}
    \caption{Total anomalous tunneling current $I_{\rm int}+I_{{\rm A}}$ contributing to the third expansion order as a function of (a) chemical potential bias $\Delta\mu/\epsilon_{\rm F}$ and (b) momentum $k/k_{\rm F}$ with different momentum cutoff $\Lambda$. Again we take $|\Delta|/\epsilon_{\rm F}=0.2$.}
    \label{fig:Iint+IA}
\end{figure}
Since the physical anomalous response is the sum of the two pair-correlated contributions, it is useful to consider $I_{\mathrm{anom},k}=I_{\mathrm{int},k}+I_{A,k}$.
Fig.~\ref{fig:Iint+IA} shows the bias and momentum dependence of $I_{\mathrm{int},k}+I_{A,k}$, which characterizes the complete anomalous correction to QGP--2SC tunneling at order $\mathcal{T}^4$. Near equilibrium, the quasiparticle excitation of red and green quarks is suppressed by the pairing gap, and thus $I_{\rm anom}$ dominates, where the conductance of the latter affects the relaxation of density difference between QGP and 2SC phases~\cite{Supplement}.

We briefly note that the chemical potential bias $\Delta\mu$ leads to the quark transfer between QGP and 2SC phases and at the same time raises a possibility of color charge transfer.
In the Supplemental Material~\cite{Supplement},
we discuss details of color charge tunneling transport.

\textit{Conclusion}.---In this work, we develop a Schwinger-Keldysh formalism for effective quantum field theory to study the transport through a QGP--2SC interface. Pairing between red and green quarks generates the pair-correlated transport including an intermediate current and Andreev reflection. The latter represents the conversion of a QGP quark into a QGP hole accompanied by Cooper-pair injection into the 2SC condensate. We investigate the bias- and momentum- dependence of the tunneling currents. Both anomalous contributions are enhanced when the bias is comparable to the 2SC gap, and their momentum dependence is governed by the resonance condition $k\simeq\mu_{\rm QGP}-\Delta \mu\pm|\Delta|$.

Our work provides a microscopic description of transport phenomena at interfaces involving color-superconducting quark matters. Future extensions of this work may include quantitative studies on imbalance relaxation towards equilibrium and the color-transfer-induced restoring force. Understanding these physics may advance studies of nuclear astrophysics, particularly in connection with compact-star cooling processes and glitch activities.

\begin{acknowledgments} 
T.~Z. acknowledges support from HK GRF (Grant No.~17306024), CRF (Grants No.~C6009-20G, No.~C7012-21G, No.~C4050-23GF), CRS-HKU701/24 and a RGC Fellowship Award No.~HKU~RFS2223-7S03.
This work is supported by JSPS KAKENHI for Grants
(Nos.~JP22K13981, JP23K22429, JP26K07063, JP26K07107) from MEXT, Japan.
\end{acknowledgments}


\bibliographystyle{apsrev4-2}
\bibliography{ref.bib}

\begin{thebibliography}{47}%
\makeatletter
\providecommand \@ifxundefined [1]{%
 \@ifx{#1\undefined}
}%
\providecommand \@ifnum [1]{%
 \ifnum #1\expandafter \@firstoftwo
 \else \expandafter \@secondoftwo
 \fi
}%
\providecommand \@ifx [1]{%
 \ifx #1\expandafter \@firstoftwo
 \else \expandafter \@secondoftwo
 \fi
}%
\providecommand \natexlab [1]{#1}%
\providecommand \enquote  [1]{``#1''}%
\providecommand \bibnamefont  [1]{#1}%
\providecommand \bibfnamefont [1]{#1}%
\providecommand \citenamefont [1]{#1}%
\providecommand \href@noop [0]{\@secondoftwo}%
\providecommand \href [0]{\begingroup \@sanitize@url \@href}%
\providecommand \@href[1]{\@@startlink{#1}\@@href}%
\providecommand \@@href[1]{\endgroup#1\@@endlink}%
\providecommand \@sanitize@url [0]{\catcode `\\12\catcode `\$12\catcode `\&12\catcode `\#12\catcode `\^12\catcode `\_12\catcode `\%12\relax}%
\providecommand \@@startlink[1]{}%
\providecommand \@@endlink[0]{}%
\providecommand \url  [0]{\begingroup\@sanitize@url \@url }%
\providecommand \@url [1]{\endgroup\@href {#1}{\urlprefix }}%
\providecommand \urlprefix  [0]{URL }%
\providecommand \Eprint [0]{\href }%
\providecommand \doibase [0]{https://doi.org/}%
\providecommand \selectlanguage [0]{\@gobble}%
\providecommand \bibinfo  [0]{\@secondoftwo}%
\providecommand \bibfield  [0]{\@secondoftwo}%
\providecommand \translation [1]{[#1]}%
\providecommand \BibitemOpen [0]{}%
\providecommand \bibitemStop [0]{}%
\providecommand \bibitemNoStop [0]{.\EOS\space}%
\providecommand \EOS [0]{\spacefactor3000\relax}%
\providecommand \BibitemShut  [1]{\csname bibitem#1\endcsname}%
\let\auto@bib@innerbib\@empty
\bibitem [{\citenamefont {Dean}\ and\ \citenamefont {Hjorth-Jensen}(2003)}]{RevModPhys.75.607}%
  \BibitemOpen
  \bibfield  {author} {\bibinfo {author} {\bibfnamefont {D.~J.}\ \bibnamefont {Dean}}\ and\ \bibinfo {author} {\bibfnamefont {M.}~\bibnamefont {Hjorth-Jensen}},\ }\href {https://doi.org/10.1103/RevModPhys.75.607} {\bibfield  {journal} {\bibinfo  {journal} {Rev. Mod. Phys.}\ }\textbf {\bibinfo {volume} {75}},\ \bibinfo {pages} {607} (\bibinfo {year} {2003})}\BibitemShut {NoStop}%
\bibitem [{\citenamefont {Baym}\ \emph {et~al.}(2018)\citenamefont {Baym}, \citenamefont {Hatsuda}, \citenamefont {Kojo}, \citenamefont {Powell}, \citenamefont {Song},\ and\ \citenamefont {Takatsuka}}]{Baym_2018}%
  \BibitemOpen
  \bibfield  {author} {\bibinfo {author} {\bibfnamefont {G.}~\bibnamefont {Baym}}, \bibinfo {author} {\bibfnamefont {T.}~\bibnamefont {Hatsuda}}, \bibinfo {author} {\bibfnamefont {T.}~\bibnamefont {Kojo}}, \bibinfo {author} {\bibfnamefont {P.~D.}\ \bibnamefont {Powell}}, \bibinfo {author} {\bibfnamefont {Y.}~\bibnamefont {Song}},\ and\ \bibinfo {author} {\bibfnamefont {T.}~\bibnamefont {Takatsuka}},\ }\href {https://doi.org/10.1088/1361-6633/aaae14} {\bibfield  {journal} {\bibinfo  {journal} {Reports on Progress in Physics}\ }\textbf {\bibinfo {volume} {81}},\ \bibinfo {pages} {056902} (\bibinfo {year} {2018})}\BibitemShut {NoStop}%
\bibitem [{\citenamefont {Haskell}\ and\ \citenamefont {Sedrakian}(2018)}]{haskell2018superfluidity}%
  \BibitemOpen
  \bibfield  {author} {\bibinfo {author} {\bibfnamefont {B.}~\bibnamefont {Haskell}}\ and\ \bibinfo {author} {\bibfnamefont {A.}~\bibnamefont {Sedrakian}},\ }\href@noop {} {\bibfield  {journal} {\bibinfo  {journal} {The Physics and Astrophysics of Neutron Stars}\ ,\ \bibinfo {pages} {401}} (\bibinfo {year} {2018})}\BibitemShut {NoStop}%
\bibitem [{\citenamefont {Barrois}(1977)}]{BARROIS1977390}%
  \BibitemOpen
  \bibfield  {author} {\bibinfo {author} {\bibfnamefont {B.~C.}\ \bibnamefont {Barrois}},\ }\href {https://doi.org/https://doi.org/10.1016/0550-3213(77)90123-7} {\bibfield  {journal} {\bibinfo  {journal} {Nuclear Physics B}\ }\textbf {\bibinfo {volume} {129}},\ \bibinfo {pages} {390} (\bibinfo {year} {1977})}\BibitemShut {NoStop}%
\bibitem [{\citenamefont {Frautschi}(1980)}]{frautschi1980asymptotic}%
  \BibitemOpen
  \bibfield  {author} {\bibinfo {author} {\bibfnamefont {S.~C.}\ \bibnamefont {Frautschi}},\ }\href@noop {} {\bibfield  {journal} {\bibinfo  {journal} {Hadronic matter at extreme energy density}\ ,\ \bibinfo {pages} {19}} (\bibinfo {year} {1980})}\BibitemShut {NoStop}%
\bibitem [{\citenamefont {Alford}\ \emph {et~al.}(2008)\citenamefont {Alford}, \citenamefont {Schmitt}, \citenamefont {Rajagopal},\ and\ \citenamefont {Sch\"afer}}]{RevModPhys.80.1455}%
  \BibitemOpen
  \bibfield  {author} {\bibinfo {author} {\bibfnamefont {M.~G.}\ \bibnamefont {Alford}}, \bibinfo {author} {\bibfnamefont {A.}~\bibnamefont {Schmitt}}, \bibinfo {author} {\bibfnamefont {K.}~\bibnamefont {Rajagopal}},\ and\ \bibinfo {author} {\bibfnamefont {T.}~\bibnamefont {Sch\"afer}},\ }\href {https://doi.org/10.1103/RevModPhys.80.1455} {\bibfield  {journal} {\bibinfo  {journal} {Rev. Mod. Phys.}\ }\textbf {\bibinfo {volume} {80}},\ \bibinfo {pages} {1455} (\bibinfo {year} {2008})}\BibitemShut {NoStop}%
\bibitem [{\citenamefont {Andreev}(1964)}]{osti_4071988}%
  \BibitemOpen
  \bibfield  {author} {\bibinfo {author} {\bibfnamefont {A.~F.}\ \bibnamefont {Andreev}},\ }\bibfield  {journal} {\bibinfo  {journal} {Zh. Eksperim. i Teor. Fiz.}\ }\textbf {\bibinfo {volume} {46}},\ \href {https://www.osti.gov/biblio/4071988} {} (\bibinfo {year} {1964})\BibitemShut {NoStop}%
\bibitem [{\citenamefont {Beenakker}(2008)}]{RevModPhys.80.1337}%
  \BibitemOpen
  \bibfield  {author} {\bibinfo {author} {\bibfnamefont {C.~W.~J.}\ \bibnamefont {Beenakker}},\ }\href {https://doi.org/10.1103/RevModPhys.80.1337} {\bibfield  {journal} {\bibinfo  {journal} {Rev. Mod. Phys.}\ }\textbf {\bibinfo {volume} {80}},\ \bibinfo {pages} {1337} (\bibinfo {year} {2008})}\BibitemShut {NoStop}%
\bibitem [{\citenamefont {Daley}\ \emph {et~al.}(2008)\citenamefont {Daley}, \citenamefont {Zoller},\ and\ \citenamefont {Trauzettel}}]{PhysRevLett.100.110404}%
  \BibitemOpen
  \bibfield  {author} {\bibinfo {author} {\bibfnamefont {A.~J.}\ \bibnamefont {Daley}}, \bibinfo {author} {\bibfnamefont {P.}~\bibnamefont {Zoller}},\ and\ \bibinfo {author} {\bibfnamefont {B.}~\bibnamefont {Trauzettel}},\ }\href {https://doi.org/10.1103/PhysRevLett.100.110404} {\bibfield  {journal} {\bibinfo  {journal} {Phys. Rev. Lett.}\ }\textbf {\bibinfo {volume} {100}},\ \bibinfo {pages} {110404} (\bibinfo {year} {2008})}\BibitemShut {NoStop}%
\bibitem [{\citenamefont {Zapata}\ and\ \citenamefont {Sols}(2009)}]{PhysRevLett.102.180405}%
  \BibitemOpen
  \bibfield  {author} {\bibinfo {author} {\bibfnamefont {I.}~\bibnamefont {Zapata}}\ and\ \bibinfo {author} {\bibfnamefont {F.}~\bibnamefont {Sols}},\ }\href {https://doi.org/10.1103/PhysRevLett.102.180405} {\bibfield  {journal} {\bibinfo  {journal} {Phys. Rev. Lett.}\ }\textbf {\bibinfo {volume} {102}},\ \bibinfo {pages} {180405} (\bibinfo {year} {2009})}\BibitemShut {NoStop}%
\bibitem [{\citenamefont {Zhang}\ and\ \citenamefont {Sommer}(2022)}]{PhysRevResearch.4.023231}%
  \BibitemOpen
  \bibfield  {author} {\bibinfo {author} {\bibfnamefont {D.}~\bibnamefont {Zhang}}\ and\ \bibinfo {author} {\bibfnamefont {A.~T.}\ \bibnamefont {Sommer}},\ }\href {https://doi.org/10.1103/PhysRevResearch.4.023231} {\bibfield  {journal} {\bibinfo  {journal} {Phys. Rev. Res.}\ }\textbf {\bibinfo {volume} {4}},\ \bibinfo {pages} {023231} (\bibinfo {year} {2022})}\BibitemShut {NoStop}%
\bibitem [{\citenamefont {Zhang}\ \emph {et~al.}(2023{\natexlab{a}})\citenamefont {Zhang}, \citenamefont {Tajima}, \citenamefont {Sekino}, \citenamefont {Uchino},\ and\ \citenamefont {Liang}}]{zhang2023dominant}%
  \BibitemOpen
  \bibfield  {author} {\bibinfo {author} {\bibfnamefont {T.}~\bibnamefont {Zhang}}, \bibinfo {author} {\bibfnamefont {H.}~\bibnamefont {Tajima}}, \bibinfo {author} {\bibfnamefont {Y.}~\bibnamefont {Sekino}}, \bibinfo {author} {\bibfnamefont {S.}~\bibnamefont {Uchino}},\ and\ \bibinfo {author} {\bibfnamefont {H.}~\bibnamefont {Liang}},\ }\href {https://doi.org/10.1038/s42005-023-01199-9} {\bibfield  {journal} {\bibinfo  {journal} {Communications Physics}\ }\textbf {\bibinfo {volume} {6}},\ \bibinfo {pages} {86} (\bibinfo {year} {2023}{\natexlab{a}})}\BibitemShut {NoStop}%
\bibitem [{\citenamefont {Sadzikowski}(2002)}]{Sadzikowski:2002ib}%
  \BibitemOpen
  \bibfield  {author} {\bibinfo {author} {\bibfnamefont {M.}~\bibnamefont {Sadzikowski}},\ }\href@noop {} {\bibfield  {journal} {\bibinfo  {journal} {Acta Phys. Polon. B}\ }\textbf {\bibinfo {volume} {33}},\ \bibinfo {pages} {1601} (\bibinfo {year} {2002})},\ \Eprint {https://arxiv.org/abs/hep-ph/0201212} {arXiv:hep-ph/0201212} \BibitemShut {NoStop}%
\bibitem [{\citenamefont {Sadzikowski}\ and\ \citenamefont {Tachibana}(2002{\natexlab{a}})}]{PhysRevD.66.045024}%
  \BibitemOpen
  \bibfield  {author} {\bibinfo {author} {\bibfnamefont {M.}~\bibnamefont {Sadzikowski}}\ and\ \bibinfo {author} {\bibfnamefont {M.}~\bibnamefont {Tachibana}},\ }\href {https://doi.org/10.1103/PhysRevD.66.045024} {\bibfield  {journal} {\bibinfo  {journal} {Phys. Rev. D}\ }\textbf {\bibinfo {volume} {66}},\ \bibinfo {pages} {045024} (\bibinfo {year} {2002}{\natexlab{a}})}\BibitemShut {NoStop}%
\bibitem [{\citenamefont {Sadzikowski}\ and\ \citenamefont {Tachibana}(2002{\natexlab{b}})}]{sadzikowski2002andreev}%
  \BibitemOpen
  \bibfield  {author} {\bibinfo {author} {\bibfnamefont {M.}~\bibnamefont {Sadzikowski}}\ and\ \bibinfo {author} {\bibfnamefont {M.}~\bibnamefont {Tachibana}},\ }\href@noop {} {\bibfield  {journal} {\bibinfo  {journal} {Acta Phys. Polon. B}\ }\textbf {\bibinfo {volume} {33}},\ \bibinfo {pages} {4141} (\bibinfo {year} {2002}{\natexlab{b}})}\BibitemShut {NoStop}%
\bibitem [{\citenamefont {Schwinger}(1961)}]{schwinger1961brownian}%
  \BibitemOpen
  \bibfield  {author} {\bibinfo {author} {\bibfnamefont {J.}~\bibnamefont {Schwinger}},\ }\href@noop {} {\bibfield  {journal} {\bibinfo  {journal} {J. Math. Phys.}\ }\textbf {\bibinfo {volume} {2}},\ \bibinfo {pages} {407} (\bibinfo {year} {1961})}\BibitemShut {NoStop}%
\bibitem [{\citenamefont {Keldysh}(1964)}]{Keldysh}%
  \BibitemOpen
  \bibfield  {author} {\bibinfo {author} {\bibfnamefont {L.~V.}\ \bibnamefont {Keldysh}},\ }\href@noop {} {\bibfield  {journal} {\bibinfo  {journal} {Zh. Eksp. Teor. Fiz.}\ }\textbf {\bibinfo {volume} {47}},\ \bibinfo {pages} {1515} (\bibinfo {year} {1964})}\BibitemShut {NoStop}%
\bibitem [{\citenamefont {Baym}\ and\ \citenamefont {Kadanoff}(1961)}]{PhysRev.124.287}%
  \BibitemOpen
  \bibfield  {author} {\bibinfo {author} {\bibfnamefont {G.}~\bibnamefont {Baym}}\ and\ \bibinfo {author} {\bibfnamefont {L.~P.}\ \bibnamefont {Kadanoff}},\ }\href {https://doi.org/10.1103/PhysRev.124.287} {\bibfield  {journal} {\bibinfo  {journal} {Phys. Rev.}\ }\textbf {\bibinfo {volume} {124}},\ \bibinfo {pages} {287} (\bibinfo {year} {1961})}\BibitemShut {NoStop}%
\bibitem [{\citenamefont {Baym}(1962)}]{PhysRev.127.1391}%
  \BibitemOpen
  \bibfield  {author} {\bibinfo {author} {\bibfnamefont {G.}~\bibnamefont {Baym}},\ }\href {https://doi.org/10.1103/PhysRev.127.1391} {\bibfield  {journal} {\bibinfo  {journal} {Phys. Rev.}\ }\textbf {\bibinfo {volume} {127}},\ \bibinfo {pages} {1391} (\bibinfo {year} {1962})}\BibitemShut {NoStop}%
\bibitem [{\citenamefont {Rammer}(2007)}]{Rammer_2007}%
  \BibitemOpen
  \bibfield  {author} {\bibinfo {author} {\bibfnamefont {J.}~\bibnamefont {Rammer}},\ }\href@noop {} {\emph {\bibinfo {title} {Quantum Field Theory of Non-equilibrium States}}}\ (\bibinfo  {publisher} {Cambridge University Press},\ \bibinfo {year} {2007})\BibitemShut {NoStop}%
\bibitem [{\citenamefont {Stefanucci}\ and\ \citenamefont {van Leeuwen}(2013)}]{Stefanucci}%
  \BibitemOpen
  \bibfield  {author} {\bibinfo {author} {\bibfnamefont {G.}~\bibnamefont {Stefanucci}}\ and\ \bibinfo {author} {\bibfnamefont {R.}~\bibnamefont {van Leeuwen}},\ }\href@noop {} {\emph {\bibinfo {title} {Nonequilibrium Many-Body Theory of Quantum Systems: A Modern Introduction}}}\ (\bibinfo  {publisher} {Cambridge University Press},\ \bibinfo {year} {2013})\BibitemShut {NoStop}%
\bibitem [{\citenamefont {Rammer}\ and\ \citenamefont {Smith}(1986)}]{RevModPhys.58.323}%
  \BibitemOpen
  \bibfield  {author} {\bibinfo {author} {\bibfnamefont {J.}~\bibnamefont {Rammer}}\ and\ \bibinfo {author} {\bibfnamefont {H.}~\bibnamefont {Smith}},\ }\href {https://doi.org/10.1103/RevModPhys.58.323} {\bibfield  {journal} {\bibinfo  {journal} {Rev. Mod. Phys.}\ }\textbf {\bibinfo {volume} {58}},\ \bibinfo {pages} {323} (\bibinfo {year} {1986})}\BibitemShut {NoStop}%
\bibitem [{\citenamefont {Jauho}\ \emph {et~al.}(1994)\citenamefont {Jauho}, \citenamefont {Wingreen},\ and\ \citenamefont {Meir}}]{PhysRevB.50.5528}%
  \BibitemOpen
  \bibfield  {author} {\bibinfo {author} {\bibfnamefont {A.-P.}\ \bibnamefont {Jauho}}, \bibinfo {author} {\bibfnamefont {N.~S.}\ \bibnamefont {Wingreen}},\ and\ \bibinfo {author} {\bibfnamefont {Y.}~\bibnamefont {Meir}},\ }\href {https://doi.org/10.1103/PhysRevB.50.5528} {\bibfield  {journal} {\bibinfo  {journal} {Phys. Rev. B}\ }\textbf {\bibinfo {volume} {50}},\ \bibinfo {pages} {5528} (\bibinfo {year} {1994})}\BibitemShut {NoStop}%
\bibitem [{\citenamefont {Meir}\ and\ \citenamefont {Wingreen}(1992)}]{PhysRevLett.68.2512}%
  \BibitemOpen
  \bibfield  {author} {\bibinfo {author} {\bibfnamefont {Y.}~\bibnamefont {Meir}}\ and\ \bibinfo {author} {\bibfnamefont {N.~S.}\ \bibnamefont {Wingreen}},\ }\href {https://doi.org/10.1103/PhysRevLett.68.2512} {\bibfield  {journal} {\bibinfo  {journal} {Phys. Rev. Lett.}\ }\textbf {\bibinfo {volume} {68}},\ \bibinfo {pages} {2512} (\bibinfo {year} {1992})}\BibitemShut {NoStop}%
\bibitem [{\citenamefont {Belzig}\ \emph {et~al.}(1999)\citenamefont {Belzig}, \citenamefont {Wilhelm}, \citenamefont {Bruder}, \citenamefont {Schön},\ and\ \citenamefont {Zaikin}}]{BELZIG19991251}%
  \BibitemOpen
  \bibfield  {author} {\bibinfo {author} {\bibfnamefont {W.}~\bibnamefont {Belzig}}, \bibinfo {author} {\bibfnamefont {F.~K.}\ \bibnamefont {Wilhelm}}, \bibinfo {author} {\bibfnamefont {C.}~\bibnamefont {Bruder}}, \bibinfo {author} {\bibfnamefont {G.}~\bibnamefont {Schön}},\ and\ \bibinfo {author} {\bibfnamefont {A.~D.}\ \bibnamefont {Zaikin}},\ }\href {https://doi.org/https://doi.org/10.1006/spmi.1999.0710} {\bibfield  {journal} {\bibinfo  {journal} {Superlattices and Microstructures}\ }\textbf {\bibinfo {volume} {25}},\ \bibinfo {pages} {1251} (\bibinfo {year} {1999})}\BibitemShut {NoStop}%
\bibitem [{\citenamefont {Wang}\ \emph {et~al.}(2014)\citenamefont {Wang}, \citenamefont {Agarwalla}, \citenamefont {Li},\ and\ \citenamefont {Thingna}}]{wang2014nonequilibrium}%
  \BibitemOpen
  \bibfield  {author} {\bibinfo {author} {\bibfnamefont {J.-S.}\ \bibnamefont {Wang}}, \bibinfo {author} {\bibfnamefont {B.~K.}\ \bibnamefont {Agarwalla}}, \bibinfo {author} {\bibfnamefont {H.}~\bibnamefont {Li}},\ and\ \bibinfo {author} {\bibfnamefont {J.}~\bibnamefont {Thingna}},\ }\href {https://doi.org/10.1007/s11467-013-0340-x} {\bibfield  {journal} {\bibinfo  {journal} {Frontiers of Physics}\ }\textbf {\bibinfo {volume} {9}},\ \bibinfo {pages} {673} (\bibinfo {year} {2014})}\BibitemShut {NoStop}%
\bibitem [{\citenamefont {Matsuo}\ \emph {et~al.}(2018)\citenamefont {Matsuo}, \citenamefont {Ohnuma}, \citenamefont {Kato},\ and\ \citenamefont {Maekawa}}]{PhysRevLett.120.037201}%
  \BibitemOpen
  \bibfield  {author} {\bibinfo {author} {\bibfnamefont {M.}~\bibnamefont {Matsuo}}, \bibinfo {author} {\bibfnamefont {Y.}~\bibnamefont {Ohnuma}}, \bibinfo {author} {\bibfnamefont {T.}~\bibnamefont {Kato}},\ and\ \bibinfo {author} {\bibfnamefont {S.}~\bibnamefont {Maekawa}},\ }\href {https://doi.org/10.1103/PhysRevLett.120.037201} {\bibfield  {journal} {\bibinfo  {journal} {Phys. Rev. Lett.}\ }\textbf {\bibinfo {volume} {120}},\ \bibinfo {pages} {037201} (\bibinfo {year} {2018})}\BibitemShut {NoStop}%
\bibitem [{\citenamefont {Kato}\ \emph {et~al.}(2019)\citenamefont {Kato}, \citenamefont {Ohnuma}, \citenamefont {Matsuo}, \citenamefont {Rech}, \citenamefont {Jonckheere},\ and\ \citenamefont {Martin}}]{PhysRevB.99.144411}%
  \BibitemOpen
  \bibfield  {author} {\bibinfo {author} {\bibfnamefont {T.}~\bibnamefont {Kato}}, \bibinfo {author} {\bibfnamefont {Y.}~\bibnamefont {Ohnuma}}, \bibinfo {author} {\bibfnamefont {M.}~\bibnamefont {Matsuo}}, \bibinfo {author} {\bibfnamefont {J.}~\bibnamefont {Rech}}, \bibinfo {author} {\bibfnamefont {T.}~\bibnamefont {Jonckheere}},\ and\ \bibinfo {author} {\bibfnamefont {T.}~\bibnamefont {Martin}},\ }\href {https://doi.org/10.1103/PhysRevB.99.144411} {\bibfield  {journal} {\bibinfo  {journal} {Phys. Rev. B}\ }\textbf {\bibinfo {volume} {99}},\ \bibinfo {pages} {144411} (\bibinfo {year} {2019})}\BibitemShut {NoStop}%
\bibitem [{\citenamefont {Ominato}\ \emph {et~al.}(2020)\citenamefont {Ominato}, \citenamefont {Fujimoto},\ and\ \citenamefont {Matsuo}}]{PhysRevLett.124.166803}%
  \BibitemOpen
  \bibfield  {author} {\bibinfo {author} {\bibfnamefont {Y.}~\bibnamefont {Ominato}}, \bibinfo {author} {\bibfnamefont {J.}~\bibnamefont {Fujimoto}},\ and\ \bibinfo {author} {\bibfnamefont {M.}~\bibnamefont {Matsuo}},\ }\href {https://doi.org/10.1103/PhysRevLett.124.166803} {\bibfield  {journal} {\bibinfo  {journal} {Phys. Rev. Lett.}\ }\textbf {\bibinfo {volume} {124}},\ \bibinfo {pages} {166803} (\bibinfo {year} {2020})}\BibitemShut {NoStop}%
\bibitem [{\citenamefont {Kawamura}\ and\ \citenamefont {Kato}(2026)}]{kawamura2026nonequilibrium}%
  \BibitemOpen
  \bibfield  {author} {\bibinfo {author} {\bibfnamefont {T.}~\bibnamefont {Kawamura}}\ and\ \bibinfo {author} {\bibfnamefont {Y.}~\bibnamefont {Kato}},\ }\href@noop {} {\bibfield  {journal} {\bibinfo  {journal} {J. Phys. Soc. Jpn.}\ }\textbf {\bibinfo {volume} {95}},\ \bibinfo {pages} {034702} (\bibinfo {year} {2026})}\BibitemShut {NoStop}%
\bibitem [{\citenamefont {Sieberer}\ \emph {et~al.}(2016)\citenamefont {Sieberer}, \citenamefont {Buchhold},\ and\ \citenamefont {Diehl}}]{Sieberer_2016}%
  \BibitemOpen
  \bibfield  {author} {\bibinfo {author} {\bibfnamefont {L.~M.}\ \bibnamefont {Sieberer}}, \bibinfo {author} {\bibfnamefont {M.}~\bibnamefont {Buchhold}},\ and\ \bibinfo {author} {\bibfnamefont {S.}~\bibnamefont {Diehl}},\ }\href {https://doi.org/10.1088/0034-4885/79/9/096001} {\bibfield  {journal} {\bibinfo  {journal} {Reports on Progress in Physics}\ }\textbf {\bibinfo {volume} {79}},\ \bibinfo {pages} {096001} (\bibinfo {year} {2016})}\BibitemShut {NoStop}%
\bibitem [{\citenamefont {Liu}\ \emph {et~al.}(2017)\citenamefont {Liu}, \citenamefont {Zhai},\ and\ \citenamefont {Zhang}}]{PhysRevA.95.013623}%
  \BibitemOpen
  \bibfield  {author} {\bibinfo {author} {\bibfnamefont {B.}~\bibnamefont {Liu}}, \bibinfo {author} {\bibfnamefont {H.}~\bibnamefont {Zhai}},\ and\ \bibinfo {author} {\bibfnamefont {S.}~\bibnamefont {Zhang}},\ }\href {https://doi.org/10.1103/PhysRevA.95.013623} {\bibfield  {journal} {\bibinfo  {journal} {Phys. Rev. A}\ }\textbf {\bibinfo {volume} {95}},\ \bibinfo {pages} {013623} (\bibinfo {year} {2017})}\BibitemShut {NoStop}%
\bibitem [{\citenamefont {Uchino}\ and\ \citenamefont {Ueda}(2017)}]{PhysRevLett.118.105303}%
  \BibitemOpen
  \bibfield  {author} {\bibinfo {author} {\bibfnamefont {S.}~\bibnamefont {Uchino}}\ and\ \bibinfo {author} {\bibfnamefont {M.}~\bibnamefont {Ueda}},\ }\href {https://doi.org/10.1103/PhysRevLett.118.105303} {\bibfield  {journal} {\bibinfo  {journal} {Phys. Rev. Lett.}\ }\textbf {\bibinfo {volume} {118}},\ \bibinfo {pages} {105303} (\bibinfo {year} {2017})}\BibitemShut {NoStop}%
\bibitem [{\citenamefont {Furutani}\ and\ \citenamefont {Ohashi}(2020)}]{furutani2020strong}%
  \BibitemOpen
  \bibfield  {author} {\bibinfo {author} {\bibfnamefont {K.}~\bibnamefont {Furutani}}\ and\ \bibinfo {author} {\bibfnamefont {Y.}~\bibnamefont {Ohashi}},\ }\href@noop {} {\bibfield  {journal} {\bibinfo  {journal} {J. Low Temp. Phys.}\ }\textbf {\bibinfo {volume} {201}},\ \bibinfo {pages} {49} (\bibinfo {year} {2020})}\BibitemShut {NoStop}%
\bibitem [{\citenamefont {Tajima}\ \emph {et~al.}(2022)\citenamefont {Tajima}, \citenamefont {Oue},\ and\ \citenamefont {Matsuo}}]{PhysRevA.106.033310}%
  \BibitemOpen
  \bibfield  {author} {\bibinfo {author} {\bibfnamefont {H.}~\bibnamefont {Tajima}}, \bibinfo {author} {\bibfnamefont {D.}~\bibnamefont {Oue}},\ and\ \bibinfo {author} {\bibfnamefont {M.}~\bibnamefont {Matsuo}},\ }\href {https://doi.org/10.1103/PhysRevA.106.033310} {\bibfield  {journal} {\bibinfo  {journal} {Phys. Rev. A}\ }\textbf {\bibinfo {volume} {106}},\ \bibinfo {pages} {033310} (\bibinfo {year} {2022})}\BibitemShut {NoStop}%
\bibitem [{\citenamefont {Tajima}\ \emph {et~al.}(2023)\citenamefont {Tajima}, \citenamefont {Oue}, \citenamefont {Matsuo},\ and\ \citenamefont {Kato}}]{tajima2023nonequilibrium}%
  \BibitemOpen
  \bibfield  {author} {\bibinfo {author} {\bibfnamefont {H.}~\bibnamefont {Tajima}}, \bibinfo {author} {\bibfnamefont {D.}~\bibnamefont {Oue}}, \bibinfo {author} {\bibfnamefont {M.}~\bibnamefont {Matsuo}},\ and\ \bibinfo {author} {\bibfnamefont {T.}~\bibnamefont {Kato}},\ }\href@noop {} {\bibfield  {journal} {\bibinfo  {journal} {PNAS nexus}\ }\textbf {\bibinfo {volume} {2}},\ \bibinfo {pages} {pgad045} (\bibinfo {year} {2023})}\BibitemShut {NoStop}%
\bibitem [{\citenamefont {Zhang}\ \emph {et~al.}(2023{\natexlab{b}})\citenamefont {Zhang}, \citenamefont {Oue}, \citenamefont {Tajima}, \citenamefont {Matsuo},\ and\ \citenamefont {Liang}}]{PhysRevB.108.155303}%
  \BibitemOpen
  \bibfield  {author} {\bibinfo {author} {\bibfnamefont {T.}~\bibnamefont {Zhang}}, \bibinfo {author} {\bibfnamefont {D.}~\bibnamefont {Oue}}, \bibinfo {author} {\bibfnamefont {H.}~\bibnamefont {Tajima}}, \bibinfo {author} {\bibfnamefont {M.}~\bibnamefont {Matsuo}},\ and\ \bibinfo {author} {\bibfnamefont {H.}~\bibnamefont {Liang}},\ }\href {https://doi.org/10.1103/PhysRevB.108.155303} {\bibfield  {journal} {\bibinfo  {journal} {Phys. Rev. B}\ }\textbf {\bibinfo {volume} {108}},\ \bibinfo {pages} {155303} (\bibinfo {year} {2023}{\natexlab{b}})}\BibitemShut {NoStop}%
\bibitem [{\citenamefont {Zhang}\ \emph {et~al.}(2024{\natexlab{a}})\citenamefont {Zhang}, \citenamefont {Tajima},\ and\ \citenamefont {Liang}}]{PhysRevApplied.21.L031001}%
  \BibitemOpen
  \bibfield  {author} {\bibinfo {author} {\bibfnamefont {T.}~\bibnamefont {Zhang}}, \bibinfo {author} {\bibfnamefont {H.}~\bibnamefont {Tajima}},\ and\ \bibinfo {author} {\bibfnamefont {H.}~\bibnamefont {Liang}},\ }\href {https://doi.org/10.1103/PhysRevApplied.21.L031001} {\bibfield  {journal} {\bibinfo  {journal} {Phys. Rev. Appl.}\ }\textbf {\bibinfo {volume} {21}},\ \bibinfo {pages} {L031001} (\bibinfo {year} {2024}{\natexlab{a}})}\BibitemShut {NoStop}%
\bibitem [{\citenamefont {Zhang}\ \emph {et~al.}(2024{\natexlab{b}})\citenamefont {Zhang}, \citenamefont {Guo}, \citenamefont {Tajima},\ and\ \citenamefont {Liang}}]{PhysRevB.110.064512}%
  \BibitemOpen
  \bibfield  {author} {\bibinfo {author} {\bibfnamefont {T.}~\bibnamefont {Zhang}}, \bibinfo {author} {\bibfnamefont {Y.}~\bibnamefont {Guo}}, \bibinfo {author} {\bibfnamefont {H.}~\bibnamefont {Tajima}},\ and\ \bibinfo {author} {\bibfnamefont {H.}~\bibnamefont {Liang}},\ }\href {https://doi.org/10.1103/PhysRevB.110.064512} {\bibfield  {journal} {\bibinfo  {journal} {Phys. Rev. B}\ }\textbf {\bibinfo {volume} {110}},\ \bibinfo {pages} {064512} (\bibinfo {year} {2024}{\natexlab{b}})}\BibitemShut {NoStop}%
\bibitem [{\citenamefont {Kawamura}\ and\ \citenamefont {Ohashi}(2024)}]{kawamura2024non}%
  \BibitemOpen
  \bibfield  {author} {\bibinfo {author} {\bibfnamefont {T.}~\bibnamefont {Kawamura}}\ and\ \bibinfo {author} {\bibfnamefont {Y.}~\bibnamefont {Ohashi}},\ }\href@noop {} {\bibfield  {journal} {\bibinfo  {journal} {AAPPS Bulletin}\ }\textbf {\bibinfo {volume} {34}},\ \bibinfo {pages} {31} (\bibinfo {year} {2024})}\BibitemShut {NoStop}%
\bibitem [{\citenamefont {Sekino}\ \emph {et~al.}(2024)\citenamefont {Sekino}, \citenamefont {Ominato}, \citenamefont {Tajima}, \citenamefont {Uchino},\ and\ \citenamefont {Matsuo}}]{PhysRevLett.133.163402}%
  \BibitemOpen
  \bibfield  {author} {\bibinfo {author} {\bibfnamefont {Y.}~\bibnamefont {Sekino}}, \bibinfo {author} {\bibfnamefont {Y.}~\bibnamefont {Ominato}}, \bibinfo {author} {\bibfnamefont {H.}~\bibnamefont {Tajima}}, \bibinfo {author} {\bibfnamefont {S.}~\bibnamefont {Uchino}},\ and\ \bibinfo {author} {\bibfnamefont {M.}~\bibnamefont {Matsuo}},\ }\href {https://doi.org/10.1103/PhysRevLett.133.163402} {\bibfield  {journal} {\bibinfo  {journal} {Phys. Rev. Lett.}\ }\textbf {\bibinfo {volume} {133}},\ \bibinfo {pages} {163402} (\bibinfo {year} {2024})}\BibitemShut {NoStop}%
\bibitem [{\citenamefont {Zhang}\ \emph {et~al.}(2025)\citenamefont {Zhang}, \citenamefont {Tajima},\ and\ \citenamefont {Tachibana}}]{zhang2025}%
  \BibitemOpen
  \bibfield  {author} {\bibinfo {author} {\bibfnamefont {T.}~\bibnamefont {Zhang}}, \bibinfo {author} {\bibfnamefont {H.}~\bibnamefont {Tajima}},\ and\ \bibinfo {author} {\bibfnamefont {M.}~\bibnamefont {Tachibana}},\ }\href {https://doi.org/10.1103/7bj7-b15c} {\bibfield  {journal} {\bibinfo  {journal} {Phys. Rev. D}\ }\textbf {\bibinfo {volume} {112}},\ \bibinfo {pages} {123031} (\bibinfo {year} {2025})}\BibitemShut {NoStop}%
\bibitem [{\citenamefont {Steiner}\ \emph {et~al.}(2002)\citenamefont {Steiner}, \citenamefont {Reddy},\ and\ \citenamefont {Prakash}}]{PhysRevD.66.094007}%
  \BibitemOpen
  \bibfield  {author} {\bibinfo {author} {\bibfnamefont {A.~W.}\ \bibnamefont {Steiner}}, \bibinfo {author} {\bibfnamefont {S.}~\bibnamefont {Reddy}},\ and\ \bibinfo {author} {\bibfnamefont {M.}~\bibnamefont {Prakash}},\ }\href {https://doi.org/10.1103/PhysRevD.66.094007} {\bibfield  {journal} {\bibinfo  {journal} {Phys. Rev. D}\ }\textbf {\bibinfo {volume} {66}},\ \bibinfo {pages} {094007} (\bibinfo {year} {2002})}\BibitemShut {NoStop}%
\bibitem [{\citenamefont {Alford}(2001)}]{alford2001color}%
  \BibitemOpen
  \bibfield  {author} {\bibinfo {author} {\bibfnamefont {M.}~\bibnamefont {Alford}},\ }\href {https://doi.org/doi.org/10.1146/annurev.nucl.51.101701.132449} {\bibfield  {journal} {\bibinfo  {journal} {Annual Review of Nuclear and Particle Science}\ }\textbf {\bibinfo {volume} {51}},\ \bibinfo {pages} {131} (\bibinfo {year} {2001})}\BibitemShut {NoStop}%
\bibitem [{Sup()}]{Supplement}%
  \BibitemOpen
  \href@noop {} {}\bibinfo {howpublished} {See Supplemental Material for details of Green's functions in QGP and 2SC phases and quasistationary relaxation of the density change.}\BibitemShut {Stop}%
\bibitem [{\citenamefont {Iida}\ and\ \citenamefont {Fukushima}(2006)}]{PhysRevD.74.074020}%
  \BibitemOpen
  \bibfield  {author} {\bibinfo {author} {\bibfnamefont {K.}~\bibnamefont {Iida}}\ and\ \bibinfo {author} {\bibfnamefont {K.}~\bibnamefont {Fukushima}},\ }\href {https://doi.org/10.1103/PhysRevD.74.074020} {\bibfield  {journal} {\bibinfo  {journal} {Phys. Rev. D}\ }\textbf {\bibinfo {volume} {74}},\ \bibinfo {pages} {074020} (\bibinfo {year} {2006})}\BibitemShut {NoStop}%
\bibitem [{\citenamefont {Fayazbakhsh}\ and\ \citenamefont {Sadooghi}(2010)}]{PhysRevD.82.045010}%
  \BibitemOpen
  \bibfield  {author} {\bibinfo {author} {\bibfnamefont {S.}~\bibnamefont {Fayazbakhsh}}\ and\ \bibinfo {author} {\bibfnamefont {N.}~\bibnamefont {Sadooghi}},\ }\href {https://doi.org/10.1103/PhysRevD.82.045010} {\bibfield  {journal} {\bibinfo  {journal} {Phys. Rev. D}\ }\textbf {\bibinfo {volume} {82}},\ \bibinfo {pages} {045010} (\bibinfo {year} {2010})}\BibitemShut {NoStop}%
\end{thebibliography}%

\newpage

\begin{widetext}

{\Large \textbf{Supplemental Material for ``Nonequilibrium Andreev transport at the QGP-2SC interface''}}

\section{Green's functions in QGP and 2SC phases}

Introducing the Nambu-Gor'kov spinors
\begin{equation}\label{eq:NGspinors}
\begingroup
\renewcommand{\arraystretch}{1.5}
  A^i_{\rm QGP,\bm{k}a}
  =
  \begin{pmatrix}
    q^i_{\bm{k}a} \\
    Cq^{i\dagger}_{-\bm{k}a}
  \end{pmatrix},
  \qquad
  A^i_{\rm 2SC,\bm{k}a}
  =
  \begin{pmatrix}
    Q^i_{\bm{k}a} \\
    CQ^{i\dagger}_{-\bm{k}a}
  \end{pmatrix}.
\endgroup
\end{equation}
we define the contour-ordered Green's functions
\begin{align}\label{eq:NGGF}
  i \mathcal{G}^{ij}_{\alpha,\bm{k}ab}(t,t')=&
  \big\langle {\rm T}_C A^i_{\alpha,\bm{k}a}(t)
  A^{j\dagger}_{\alpha,\bm{k}b}(t')\big\rangle_0,
\end{align}
with $\alpha={\rm QGP},{\rm 2SC}$. The matrix structure reads
\begin{equation}\label{eq:NGmatrix}
  i \mathcal{G}^{ij}_{\alpha,\bm{k}ab}(t,t')=
  \begin{pmatrix}
  G^{ij+}_{\alpha,\bm{k}ab}(t,t') & F^{ij-}_{\alpha,\bm{k}ab}(t,t')\\
  F^{ij+}_{\alpha,\bm{k}ab}(t,t') & G^{ij-}_{\alpha,\bm{k}ab}(t,t')
  \end{pmatrix}.
\end{equation}
In the QGP region, the off-diagonal elements $F^{ij,\pm}_{\rm QGP,\bm{k},ab}=0$ because there is no pairing condensate. Moreover, normal components are diagonal in color and flavor space, $G^{ij\pm}_{{\rm QGP},\bm{k}ab}=\delta^{ij}\delta_{ab}
G^{\pm}_{\rm QGP,\bm{k}}$.

For an explicit evaluation, we use massless relativistic propagators. The normal Green's functions are given by
\begin{align}
  G^+_{\bm{k}}(\omega)&=
  \frac{\gamma_0\Lambda^+_{\bm{k}}}{[\omega+i\eta+(\mu-k)]}
  +\frac{\gamma_0\Lambda^-_{\bm{k}}}{[\omega+i\eta+(\mu+k)]},
  \label{eq:Gplus}\\
  G^-_{\bm{k}}(\omega)&=
  \frac{\gamma_0\Lambda^-_{\bm{k}}}{[\omega+i\eta-(\mu-k)]}
  +\frac{\gamma_0\Lambda^+_{\bm{k}}}{2[\omega+i\eta-(\mu+k)]}.
  \label{eq:Gminus}
\end{align}
The anomalous propagators in the 2SC phase are
\begin{align}
  F^+_{\bm{k}}(\omega)=\Delta&\gamma_5\Biggl[
  \frac{\Lambda^-_{\bm{k}}}{(\omega+i\eta)^2-(E^{+}_{\bm{k}})^2}
  +\frac{\Lambda^+_{\bm{k}}}{(\omega+i\eta)^2-(E^{-}_{\bm{k}})^2}
  \Biggr],
  \label{eq:Fplus}\\
  F^-_{\bm{k}}(\omega)=-&\Delta\gamma_5\Biggl[
  \frac{\Lambda^+_{\bm{k}}}{(\omega+i\eta)^2-(E^{+}_{\bm{k}})^2}
  +\frac{\Lambda^-_{\bm{k}}}{(\omega+i\eta)^2-(E^{+}_{\bm{k}})^2}
  \Bigg],
  \label{eq:Fminus}
\end{align}
where $E_{\bm{k}}^\pm=\sqrt{(k\mp\mu)^2+|\Delta|^2}$ are the quasiparticle excitation dispersions in the 2SC phase, corresponding to the branches associated with the positive- and negative-energy Dirac sectors. The operators
\begin{equation}\label{eq:projectors}
  \Lambda^\pm_{\bm{k}}=\frac{1}{2}\left(1\pm\gamma_0\bm{\gamma}\cdot\bm{\hat{k}}\right).
\end{equation}
project Dirac spinors onto the positive- and negative-energy branches. We note that in the isotropic-contact approximation, the angular average gives $\langle\bm{\gamma}\cdot\bm{\hat{k}}\rangle_\Omega=0$.

\section{Quasistationary relaxation}
The macroscopic charge tunneling current is given by
\begin{align}
    \mathcal{I}=\sum_{\bm{k},a,i}\langle\hat{I}_{\bm{k}ai}\rangle,
\end{align}
At a small chemical-potential bias, $\mathcal{I}$ can be rewritten as
\begin{align}
    \mathcal{I}=-\frac{d}{dt}\Delta N=G\Delta\mu+O((\Delta\mu)^2),
\end{align}
where $\Delta N\equiv (N_{\rm QGP}-N_{\rm 2SC})/2$  is the density difference between the QGP and 2Sc phases.
$G$ is a constant corresponding to the conductance in condensed matter systems.
Under the quasistationary condition,
using the thermodynamic relation
\begin{align}
    \Delta N&=\left(\frac{\partial \Delta N}{\partial \Delta\mu}\right)_T\Delta\mu\cr
    &=\frac{1}{2}\left[\left(\frac{\partial N_{\rm QGP}}{\partial\mu_{\rm QGP}}\right)_T
    +
    \left(\frac{\partial N_{\rm 2SC}}{\partial\mu_{\rm 2SC}}\right)_T
    \right]\Delta\mu\cr
    &\equiv \chi \Delta\mu,
\end{align}
we obtain the isothermal time-evolution equation
\begin{align}
    \frac{d}{dt}\Delta N=-\frac{G}{\chi}\Delta N.
\end{align}
$\chi$ corresponds to the density susceptibility associated with the equation of state.
In this sense, we obtain the relaxation of the density difference as
\begin{align}
    \Delta N(t)=\Delta N(t=0)e^{-\frac{G}{\chi}t},
\end{align}
where the relaxation time is given by $\chi/G$.
This density relaxation might be observed via the dynamical evolution of a neutron star.

\section{Comments on color transfer}

In this section, we discuss
a possibility of color charge transfer between QGP and 2SC phases. The Andreev reflection and the intermediate tunneling processes arise only for red and green color quarks, and the quasiparticle tunneling for red and green color charges are different from that for blue ones due to the modification from pairing gap. Such an imbalance in color flow creates a strong color field across the boundary surface, which eventually prevents further growth of the color charge difference. This situation is actually expected at the QGP/2SC interface because the QGP phase is color neutral whereas the 2SC phase is not. Such color forces give an effective shift on the chemical potential of quarks in the color-superconducting phase depending on the color charge: $\mu_c=\mu_3T^3+\mu_8T^8$, where
\begin{equation}
  T^3=\frac{1}{2}
  \begin{pmatrix}1&0&0\\0&-1&0\\0&0&0\end{pmatrix},
  \quad
  T^8=\frac{1}{2\sqrt{3}}
  \begin{pmatrix}1&0&0\\0&1&0\\0&0&-2\end{pmatrix}.
  \label{eq:T3T8}
\end{equation} 
The color charge tunneling current is given by
\begin{align}
    \mathcal{I}^A=-\sum_{\bm{k},a,i}\big\langle\dot{N}^{A}_{{\rm QGP},\bm{k}ai}\big\rangle=\sum_{\bm{k},a,i}\langle\hat{I}^A_{\bm{k}ai}\rangle,
\end{align}
\begin{figure}[t]
    \centering
    \includegraphics[width=12cm]{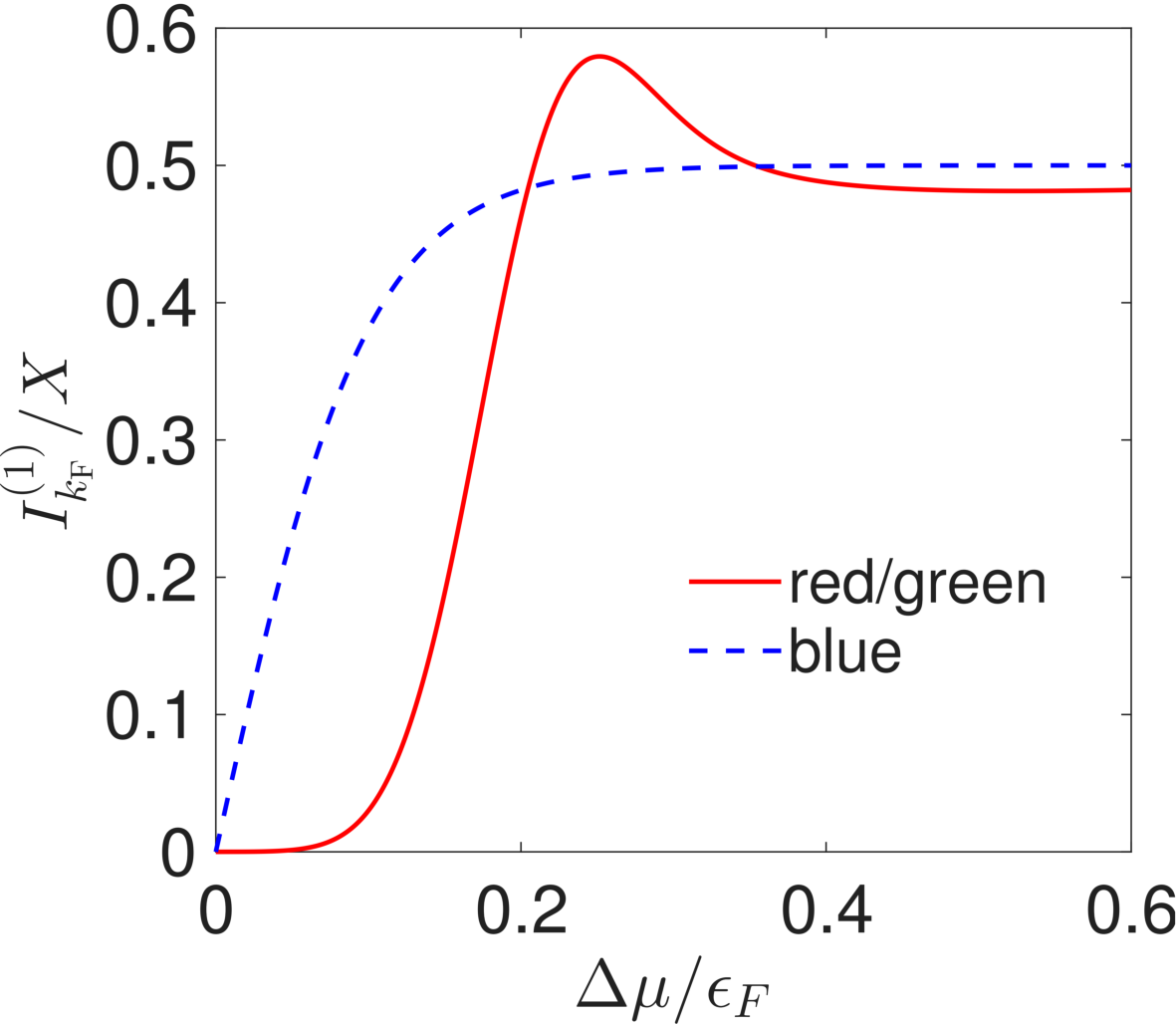}
    \caption{The leading-order quasiparticle tunneling current $I^{(1)}_{k_{\rm F}}$ of red/green and blue charges as functions of $\Delta\mu$. The value of 2sc gap is taken as $|\Delta|/\epsilon_{\rm F}=0.2$. The temperature is taken as $T/T_{\rm F}\simeq 0.05$ ($T\simeq 25$~MeV). $X=6\mathcal{T}^2k_{\rm F}^2/\pi$ is the normalizing factor.
    }
    \label{fig:Icharge}
\end{figure}
where $N^A_{\rm QGP,\bm{k}ai}=q^{i\dagger}_{\bm{k}a}T^Aq^i_{\bm{k}a}$. In The 2SC phase $\mathcal{I}^3=0$ due to the red-green symmetry, and $\mathcal{I}^8= (I_r+I_g-2 I_b)/2\sqrt{3}$. Color neutrality requires $\mathcal{I}^A=0$, leading to zero $\mu_3$ and nonzero $\mu_8$, and effective color-dependent chemical potentials: 
\begin{align}
    \mu_r=\mu_g&=\mu+\mu_8/{2\sqrt{3}},\nonumber\\
    \mu_b&=\mu-{\mu_8}/{\sqrt{3}} .
\end{align}

In general, one expects that when the system is far from equilibrium ($T, \Delta\mu \gg |\Delta|$), the transport is dominated by quasiparticle tunneling, whereas the Andreev reflection and the intermediate process for red and green charges are suppressed. At this stage, the pairing effect causes a small difference between the blue and red/green charges as shown in Fig.~\ref{fig:Icharge}, leading to a small $\mu_8$ to maintain color neutrality at the interface. Similarly, this prediction is expected to hold close to equilibrium ($T,\Delta\mu \ll |\Delta|$) because the QGP phase is color neutral, while the 2SC phase maintains a small $\mu_8$, as predicted by the bulk effective models \cite{RevModPhys.80.1455}. However, in the intermediate stage, when ($T, \Delta\mu\sim |\Delta|$), there are two competing effects. Since the red and green quasiparticles are gapped in the 2SC phase, their quasiparticle tunneling currents are suppressed, but at the same time, their anomalous current rises. In the meantime, the unpaired blue component remains intact because its quasiparticles are gapless. Consequently, the net contribution of the quasiparticle to the color current $I_8$ may become relatively large in this subgap region, leading to an effective shift of $\mu_8$ that tends to restore color neutrality. However, the quantitative predictions depend here on the dynamical details ofthe interface region, in particular onthe tunneling coupling strength $\mathcal{T}$, as well asthe structure ofthe bulk phases, and require independent study.

\end{widetext}

\end{document}